\documentclass{Interspeech}

\interspeechcameraready

\title{H-QuEST: Accelerating Query-by-Example Spoken Term Detection with Hierarchical Indexing}

\author[affiliation={1,2}]{Akanksha}{Singh}
\author[affiliation={1}]{Yi-Ping Phoebe}{Chen}
\author[affiliation={2}]{Vipul}{Arora}

\affiliation{Department of Computer Science and Information Technology}{La Trobe University}{Australia}
\affiliation{Department of Electrical Engineering}{Indian Institute of Technology Kanpur}{India}

\email{akankss20@iitk.ac.in, Phoebe.Chen@latrobe.edu.au, vipular@iitk.ac.in  }

\usepackage{comment}

\begin{document}

\maketitle

\begin{abstract}
    
        Query-by-example spoken term detection (QbE-STD) searches for matching words or phrases in an audio dataset using a sample spoken query. When annotated data is limited or unavailable, QbE-STD is often done using template matching methods like dynamic time warping (DTW), which are computationally expensive and can’t scale well. To address this, we propose H-QuEST (Hierarchical Query-by-Example Spoken Term Detection), a novel framework that accelerates spoken term retrieval by utilizing Term Frequency and Inverse Document Frequency (TF-IDF)-based sparse representations obtained through advanced audio representation learning techniques and Hierarchical Navigable Small World (HNSW) indexing with further refinement to perform search. Experimental results show that H-QuEST delivers substantial improvements in retrieval speed, without sacrificing accuracy compared to existing methods.
\end{abstract}

\section{Introduction}
Query-by-Example Spoken Term Detection (QbE-STD) is the retrieval of spoken content by matching spoken queries to audio datasets. It faces challenges in identifying spoken queries within large audio datasets due to speaker variability, environmental factors, and language-specific changes. Traditionally researchers combined Automatic Speech Recognition (ASR) systems \cite{huang2001spoken} with text-based retrieval methods to solve this task. However, these techniques often require large amounts of annotated spoken data for accurate detection. The annotation process is both tedious and demands expertise in the target language, creating a significant barrier in low-resource languages where annotated data is scarce or nonexistent. 

Studies by \cite{jansen2011efficient, park2005towards, park2007unsupervised, rasanen2015unsupervised, kamper2017embedded} show that QbE-STD is feasible even in the absence of dedicated resources. This is an alternative solution that involves pattern discovery, which identifies similarities among spoken terms directly from their acoustic feature representations. These approaches are adaptable across languages and are capable of performing tasks without needing large amounts of language-specific labeled data. 

There are two broad categories for pattern discovery: Dynamic Time Warping (DTW)-based techniques and template matching methods. In DTW-centric approaches \cite{park2005towards, zhang2010towards, gupta2011language, karthik2016fast}, the similarity between spoken terms is captured by computing the temporal alignment of acoustic features. However, DTW suffers from high computational cost, with time complexity proportional to the length of the query and audio segments, making it inefficient for large datasets. To address this, segmental DTW \cite{park2007unsupervised, dumpala2015analysis, zhang2009unsupervised} was introduced to allow for alignment at a more granular, segment-based level. Despite this improvement, segmental DTW is still sensitive to the choice of segment length and faces challenges with scalability. On the other hand, template matching techniques measure similarity between spoken terms by identifying recurring patterns within the audio. One example is using an n-gram approach \cite{rasanen2015unsupervised}, where syllable boundaries help convert variable-length segments into fixed-size representations for easier comparison. Alternatively, methods like \cite{kamper2017embedded} group fixed-size audio segments using k-means clustering, effectively identifying recurring acoustic patterns. Some researchers also integrated deep neural networks (DNNs) \cite{ram2018cnn, bhati2018unsupervised, cui2015multilingual, knill2014language, yuan2020fast} to transfer knowledge from resource-rich languages to improve performance in low-resource languages.

Recent advancements, such as the work in \cite{8462570}, have proposed end-to-end QbE-STD models using attention-based multi-hop networks that score query presence across audio segments. Additionally, \cite{9414156} uses a multi-head attention module atop a multi-layered Gated Recurrent Unit (GRU) for feature extraction and aggregation, coupled with soft-triple loss to refine performance. Although some template-matching methods involve tokenizing audio and performing brute-force TF-IDF vector searches \cite{10605547}, these still remain computationally expensive due to the high-dimensional cosine similarity matrix. Without efficient indexing methods, these systems often require linear searches across the entire dataset, which becomes computationally expensive, especially when handling short queries. Locality-sensitive hashing (LSH) \cite{gionis1999similarity} and subspace-indexing have been explored as potential solutions to improve search efficiency, but scalability remains a persistent challenge. 

In this work, we propose H-QuEST (Hierarchical Query-by-Example Spoken Term Detection), a scalable approach that combines Term Frequency and Inverse Document Frequency (TF-IDF) \cite{sparck2004idf,luhn1958automatic} vectors with Hierarchical Navigable Small World (HNSW) \cite{malkov2018efficient} indexing that refines retrieval results using the Smith-Waterman algorithm \cite{smith1981identification}. The approach leverages Wav2Vec2.0 \cite{baevski2020wav2vec} for robust feature extraction, capturing rich contextual information from audio. We demonstrate that H-QuEST improves scalability, retrieval precision, better mean average precision (MAP) and reduced query search times compared to existing approaches.

\begin{figure}[t]
    \centering
    \includegraphics[width=\linewidth]{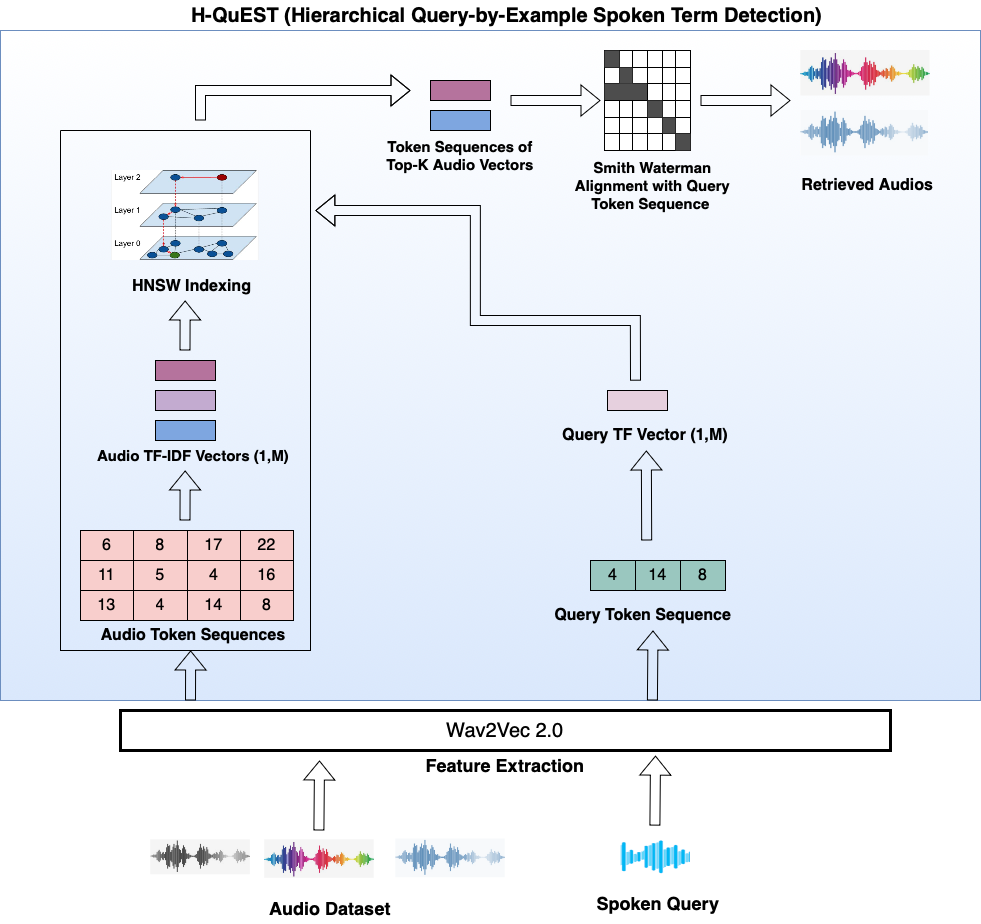}
    \caption{ Overview of the proposed H-QuEST pipeline. Audio files are first tokenized into discrete sequences using a pretrained audio tokenizer. Each sequence is transformed into a TF-IDF vector, and an HNSW graph is built over these vectors for efficient similarity search. At query time, the input audio is tokenized and vectorized in the same way, and cosine similarity is used to retrieve the top-K nearest neighbors. These candidates are then re-ranked using the Smith-Waterman algorithm for fine-grained sequence alignment and improved retrieval accuracy.}
    \label{fig:diagram}
\end{figure}

\section{Methodology}
\subsection{Audio Representation Learning}

Wav2Vec2.0 model \cite{baevski2020wav2vec}, performs audio tokenization by first extracting features from raw audio through a convolutional neural network (CNN) and then passing these features through transformer layers. The resulting feature representations capture important phonetic and acoustic information. These continuous features are clustered into discrete tokens using a process called vector quantization, where k-means clustering groups similar feature vectors and assign each cluster a unique token in the vocabulary. For our case, the token sequences of audio dataset is extracted from the last transformer layer, as it captures the most contextually rich representations. The vocabulary consists of 51 unique tokens, which are learned during the pre-training of the model on large-scale audio datasets. We define the tokenized audio sequences as:
\begin{align}
    \mathcal{D} &= \{d_i \mid 1<i \leq N\} \\
    d_i &= \{ (\tau_1, \tau_2, \ldots, \tau_n, \ldots) \mid 1 < n \leq M \} 
\end{align}
where $N$ audio files make a set $\mathcal{D}$, the token sequence of a single audio file from the dataset is portrayed by $d_i$ and $\tau_n$ stands for a token from the set of $M$ unique tokens which form the vocabulary for the audio dataset. This sequence representation enables linguistic content matching across audio files.
\subsection{Pattern Matching using HNSW Indexing}

\subsubsection{TF-IDF Vectorization}
Each token sequence $d_i$ is transformed into a vector $\mathbf{v}_i$, where each dimension corresponds to a token $\tau_n$ from the vocabulary. The value in each dimension is the Term Frequency-Inverse Document Frequency (TF-IDF) \cite{sparck2004idf,luhn1958automatic} score for that token in the sequence. 

Let $\mathcal{M} = \{\tau_1, \tau_2, \dots, \tau_M\}$ be the vocabulary of tokens, where $M$ is the size of the vocabulary. Then, the TF-IDF vector $\mathbf{v}_i$ for a sequence $d_i$ is given by:
\begin{align}
\mathbf{v}_i = \left[ \text{TF-IDF}(\tau_1, d_i), \text{TF-IDF}(\tau_2, d_i), \dots, \text{TF-IDF}(\tau_N, d_i) \right]
\end{align}

Each element in the vector $\mathbf{v}_i$ represents the importance of the corresponding token $\tau_n$ in the sequence $d_i$, calculated as:

\begin{align}
\text{TF-IDF}(\tau_n, d_i) = \text{TF}(\tau_n, d_i) \times \log \left( \frac{N}{\text{DF}(\tau_n)} \right)
\end{align}
where, $N$ is the total number of audio files in the dataset.
Thus, the vector $\mathbf{v}_i$ is a sparse representation of the audio sequence, with the components corresponding to tokens in the vocabulary and their respective TF-IDF scores. If a token does not appear in the sequence, its corresponding component in the vector is zero.
\subsubsection{Constructing the HNSW Index}
HNSW \cite{malkov2018efficient} organizes these TF-IDF \cite{sparck2004idf,luhn1958automatic} vectors in a multi-layered, navigable small-world graph, facilitating efficient nearest-neighbor search. Each audio sequence (TF-IDF vector) is inserted into a multi-layered graph: Nodes (vectors) are connected based on proximity in feature space. A higher-layer structure provides long-range links, while lower layers store local connections.
When inserting a new vector, it is connected to the closest nodes in the graph using cosine similarity:
\begin{equation}
\cos(\theta) = \frac{\vec{v_i} \cdot \vec{v_j}}{||\vec{v_i}|| ||\vec{v_j}||}
\end{equation}
where:
\begin{itemize}
    \item $\vec{v_{i}} \cdot \vec{v_j}$ is the dot product of the two TF-IDF vectors.
    \item $||\vec{v_i}||$ and $||\vec{v_j}||$ are their respective magnitudes (L2 norms).
\end{itemize}
The logarithmic search time complexity ensures quick retrieval, efficiently handling large-scale audio datasets.

Figure~\ref{fig:diagram} illustrates the steps of H-QuEST. Initially, token sequences are generated for the audio dataset, which are then used to construct an HNSW graph based on the audio TF-IDF vectors (1, M), where M represents the total number of unique tokens derived from these sequences and there are N such vectors. The query audio undergoes a similar tokenization process. Following this, a cosine similarity operation is performed between the query’s TF vector (1, M) and the nodes in the HNSW graph. This operation ranks the top-K nearest neighbour audio files to the query, which are subsequently re-ranked using the Smith-Waterman algorithm to identify the best retrieval of audio files.

\subsection{Query Search}
The query search phase begins by tokenizing the audio query using the same pre-trained Wav2Vec2.0 model \cite{baevski2020wav2vec}, to generate token sequences that are consistent with those in the audio dataset. These tokenized query sequences are represented as Term Frequency (TF) vectors, ensuring they align with the same vector space as the audio dataset.

Once the query is represented as a TF vector, the HNSW algorithm is employed to efficiently find the nearest neighbors in the dataset. The search starts from the top level of the hierarchical graph, progressively moving toward finer levels. This hierarchical structure enables a balance between global and local exploration, ensuring an efficient search process that quickly narrows down the most relevant candidates.

Cosine similarity is used to quantify the similarity between the query vector \( \vec{q} \) and an audio vector \( \vec{v_i} \) from the dataset. This metric is effective because it measures the cosine of the angle between two vectors, focusing on the direction rather than the magnitude, making it ideal for comparing tokenized audio representations. 

Indexing the audio dataset hierarchically using HNSW helps reduce the number of computations performed on negative audio files— those that do not contain the query. As the number of relevant utterances for a particular query is much smaller compared to the total number of audio files in the dataset. By focusing on the most promising candidates, the algorithm efficiently filters out irrelevant files, reducing computational overhead.
After retrieving the top-\( K \) nearest neighbour audio files based on cosine similarity scores, the results are further refined using Smith-Waterman sequence alignment. The Smith-Waterman similarity score between two token sequences \( d_1 \) and \( d_2 \) is computed using dynamic programming:
\begin{align}
H(i, j) = \max \begin{cases}
    H(i-1, j-1) + s(i, j), & \text{(match/mismatch)} \\
    H(i-1, j) - g, & \text{(deletion)} \\
    H(i, j-1) - g, & \text{(insertion)} \\
    0, & \text{(termination)}
\end{cases}
\end{align}

Where: \( H(i, j) \) represents the similarity score at position \( (i, j) \).
\( s(i, j) \) is the match/mismatch score for aligning tokens at positions \( i \) and \( j \), we use a match score of \( +2 \), a mismatch penalty of \( -1 \), and a gap penalty \( g \) of \( -2 \) for insertions and deletions.

\section{Experiments}

\subsection{Audio datasets}

\begin{itemize}
  
    \item LibriSpeech Dataset: The LibriSpeech dataset \cite{panayotov2015librispeech}, derived from LibriVox's public domain audiobooks, is used to compare H-QuEST against QbE-STD Using Attention-Based Multi-Hop Networks \cite{8462570}. The evaluation is conducted using three distinct test sets, all sourced from the train-other-500 subset of LibriSpeech: All the three testing sets are similar to those used in \cite{8462570}. From 30 unique spoken queries we get 1,500 query-segment pairs, with every query having approximately 50 examples.
    
    \item Timit Dataset: The TIMIT dataset \cite{garofolo1993timit} is split into three sets: the development set (1000 files, 50 minutes), the database set (4500 files, 3.8 hours), and the test set (800 files). The test set consists of 800 files, selected using 30 queries of varying lengths. There are 5 relevant occurrences for each query in the dataset. This is used to compare H-QuEST with An Efficient TF-IDF Based Query by Example Spoken Term Detection \cite{10605547}, which employs a brute force search using cosine similarity on TF-IDF vectors of both the query and the audio dataset. 

    \item Hey-Snips Dataset: The Hey-Snips (HS) dataset \cite{coucke2019efficient} is used to compare H-QuEST against Query-by-Example Keyword Spotting System Using Multi-Head Attention and Soft-triple Loss \cite{9414156}, the HS dataset contains instances of the keyword "Hey Snips" alongside general sentences. During comparison, three utterances are randomly selected as enrollment samples for each speaker, which are similar to the ones used in \cite{9414156} there are 40 speakers (test) with positive keyword examples and 20,543 (test) negative samples (general sentences). 10 dB babble noise is added to all test utterances. The False Rejection Rate (FRR) is computed at the utterance level for this dataset.
\end{itemize}

\begin{table*}[ht]
    \centering
    \caption{Evaluation metrics for performance analysis of  QbE-STD using TF-IDF \cite{10605547} and the proposed approach H-QuEST on TIMIT dataset across segment sizes (0.8s, 1s, 1.2s) for Precision@K (K=1,3,5) and MAP (higher is better).}
    \label{tab:brute}
    \resizebox{\textwidth}{!}{
    \begin{tabular}{|c|ccc|ccc|ccc|ccc|} \hline
        \textbf{Approach} & \multicolumn{3}{c|} {P@1} & \multicolumn{3}{c|}{P@3} & \multicolumn{3}{c|}{P@5} & \multicolumn{3}{c|}{MAP}\\ \cline{2-13}
        & 0.8s & 1s & 1.2s & 0.8s & 1s & 1.2s & 0.8s & 1s & 1.2s & 0.8s & 1s & 1.2s \\ \hline
        BoAW+DTW & 0.5357 & 0.733 & 0.733 & 0.4405 & 0.5667 & 0.633 & 0.3643 & 0.4400 & 0.5200 & 0.329 & 0.4570 & 0.5051 \\
        TF-IDF (Brute Force) & 0.5714 & 0.857 & 0.875 & 0.4809 & 0.576 & 0.649 & 0.3742 & 0.4441 & 0.5472 & 0.3375 & 0.4575 & 0.558 \\
        \hline
        \textbf{ H-QuEST (Ours)} & \textbf{0.6714} & \textbf{0.8846} & \textbf{0.895} & \textbf{0.5409} & \textbf{0.588} & \textbf{0.659} & \textbf{0.4472} & \textbf{0.4916} & \textbf{0.5627} & \textbf{0.4575} & \textbf{0.4833} & \textbf{0.648} \\
        \hline
    \end{tabular}
    }
\end{table*}

\begin{table}[ht]
    
    \caption{Comparison of MAP values (higher is better) across different approaches of QbE-STD Using Attention-Based Multi-Hop Networks \cite{8462570} and the proposed approach H-QuEST on Librispeech train-other-500 dataset.}
    \label{tab:map_comparison}
    \centering
    \begin{tabular}{|l|c|c|c|}
        \hline
        \textbf{Approach} & Test 1 & Test 2 & Test 3 \\
        \hline
        (A): DTW & 0.6173 & 0.5778 & 0.5678 \\
        (B): Network without Attention & 0.5935 & 0.5563 & 0.5468 \\
        \hline
        (C): Attention-based Network & & & \\
        \quad (1) NN & 0.6523 & 0.6246 & 0.5754 \\
        \quad (2) Cosine & 0.6331 & 0.6043 & 0.5746 \\
        \quad (3) NN+Cos & 0.6268 & 0.6370 & 0.5759 \\
        \hline
        (D): (A)+(C) & & & \\
        \quad (1) NN & 0.6720 & 0.6340 & 0.5868 \\
        \quad (2) Cosine & 0.6433 & 0.6002 & 0.5843 \\
        \quad (3) NN+Cos & 0.6451 & 0.6309 & 0.5808 \\
        \hline
        
        Inverted Index & 0.6285 & 0.6143 & 0.6035 \\
        TF-IDF (Brute-Force) & 0.6152 & 0.5996 & 0.5893 \\
        \textbf{H-QuEST (Ours)} & \textbf{0.7470} & \textbf{0.7250} & \textbf{0.6320} \\
        \hline
    \end{tabular}

\end{table}

\begin{table}[ht]
    
    \caption{Comparison of False Rejection Rate (FRR) (lower is better) on HS (Clean) and HS (Babble) across different approaches of Query-by-Example Keyword Spotting System Using Multi-Head Attention and Soft- triple Loss \cite{9414156} against the proposed approach H-QuEST.}
    \label{tab:hs_comparison}
    \centering
    \begin{tabular}{|l|c|c|}
        \hline
        \textbf{Approach} & Hey-Snips (Clean) & Hey-Snips (Babble) \\
        \hline
        Baseline (Small) & 12.09 & 11.70 \\
        Baseline (Large) & 11.70 & 9.58 \\
        QbE-KWS (Small) & 7.85 & 9.65 \\
        QbE-KWS (Large) & 6.67 & 7.38 \\
        \textbf{H-QuEST (Ours)} & \textbf{6.30} & \textbf{6.20} \\
        \hline
    \end{tabular}

\end{table}

\subsection{Experimental Setup}
The proposed H-QuEST framework is designed to improve the efficiency of spoken term retrieval in large audio datasets. It leverages sparse representations based on TF-IDF, which are derived from Wav2Vec2.0 \cite{baevski2020wav2vec}. These representations are then indexed using a Hierarchical Navigable Small World (HNSW) graph \cite{malkov2018efficient}, configured with an exploration factor (ef\_construction) of 150 and 16 neighbours per node. This setup facilitates fast nearest-neighbor search during the retrieval process. The exploration factor, denoted as (ef\_construction), plays a crucial role in controlling the balance between index construction time and search accuracy. Higher values allow for a more thorough search, which can enhance retrieval accuracy.
During retrieval, H-QuEST performs a fast approximate nearest-neighbor search over the HNSW graph to identify the top-K audio files. We experimented with \( K \in \{50, 100, 200\} \) and found that
\( K = 50 \) offers the best balance between recall and computational efficiency. To further refine the search results, H-QuEST applies the Smith-Waterman algorithm \cite{smith1981identification} to align the query sequence with each of the top-K retrieved candidates. This local sequence alignment step improves final accuracy by capturing fine-grained temporal similarities that sparse vector similarity alone may miss.

We also compare the performance of H-QuEST against a method in which we build an inverted index \( I \), which maps each token \( \tau_n \) to the set of audio files \( d_i \) containing it:
\begin{align}
    I: \tau_n \mapsto \{ d_i \mid \tau_n \in d_i \}
\end{align}

This inverted index allows for efficient retrieval of audio files containing specific tokens, eliminating the need for exhaustive searching through all the files. 

\section{Results and Discussion}
Table~\ref{tab:brute} compares Precision at K (P@K for K = 1, 3, 5) and Mean Average Precision (MAP) values of TF-IDF \cite{10605547}, BoAW + DTW \cite{george2014unsupervised}, and the proposed H-QuEST— on the TIMIT dataset \cite{garofolo1993timit} across segment sizes of 0.8s, 1s, and 1.2s. The BoAW + DTW approach exhibits relatively lower precision and MAP scores, reflecting its less effective retrieval of spoken terms. The TF-IDF method shows improvements over BoAW + DTW, with higher precision and MAP values across all segment sizes. However, the proposed H-QuEST approach outperforms both BoAW + DTW and TF-IDF in all metrics since it performs refinement of top-K retrieved audio sequences from HNSW, to get the best aligned audio sequences with respect to the query using Smith-Waterman algorithm. Thus, achieving the highest precision and MAP scores, indicating superior retrieval accuracy and efficiency.

Table~\ref{tab:map_comparison} presents a comparison of MAP values for different QbE-STD approaches, including Attention-Based Multi-Hop Networks \cite{8462570}, TF-IDF \cite{10605547}, and H-QuEST, evaluated across three test sets from the LibriSpeech dataset. Among the baseline methods, DTW delivers moderate performance but struggles with speech variability, such as differences in speakers and environmental conditions. A hybrid approach that combines DTW with attention-based networks improves performance by integrating sequence alignment with neural attention mechanisms, demonstrating the benefits of combining traditional and neural techniques.
The inverted index-based search avoids a full dataset scan by retrieving only audio files containing matching tokens, leading to reasonable MAP values. As it relies on exact token matches, it is slightly less effective than H-QuEST, which first identifies nearest neighbours, before refining search with Smith-Waterman algorithm. TF-IDF search, which ranks audio vectors by cosine similarity with the query vector, performs similar to the inverted index but remains computationally expensive due to its full comparison approach.
In contrast, H-QuEST significantly outperforms all baselines, achieving the highest MAP values across all test sets. This superior performance is attributed to H-QuEST’s hierarchical search mechanism, which efficiently indexes audio segments and refines matches, enabling it to surpass DTW-based, TF-IDF, and attention-based models in QbE-STD.

\begin{figure}[ht]
    \centering
    \includegraphics[width=\linewidth]{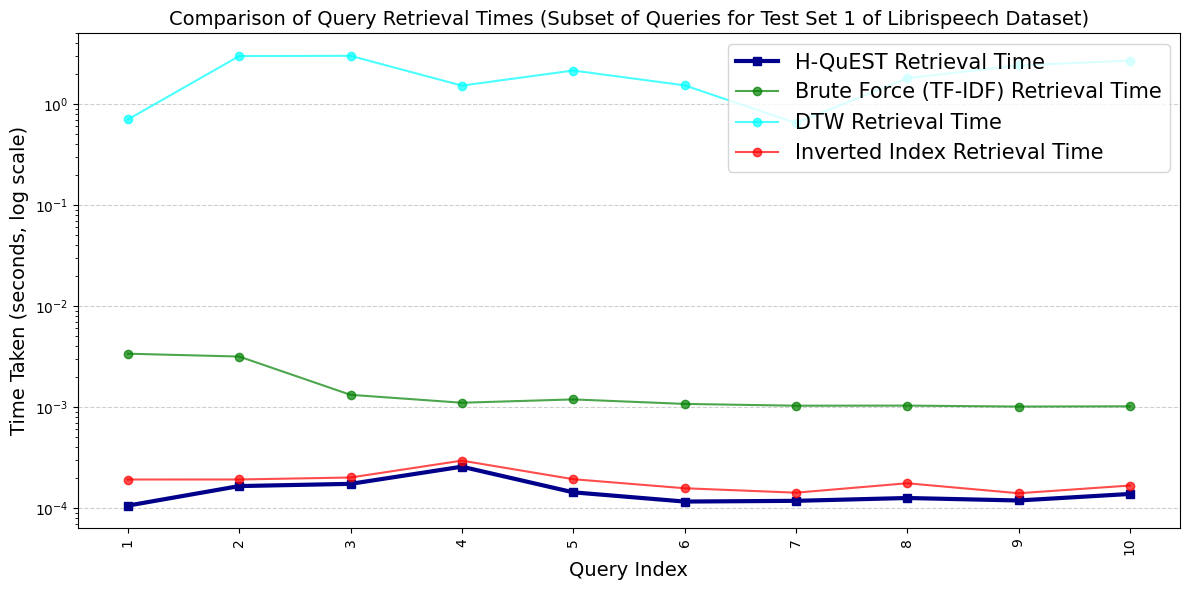}
    \caption{Comparison of query retrieval times across search methods on a Librispeech subset. The y-axis shows time (seconds, log scale), and the x-axis shows query indices. H-QuEST achieves significantly faster retrieval than DTW and remains competitive with the inverted index method.}
    \label{fig:retrieval_times}
    
\end{figure}

Figure~\ref{fig:retrieval_times} demonstrates query retrieval times for different search methods on test set 1 of the Librispeech train-other-500 dataset, with retrieval time (seconds) on a logarithmic scale. DTW exhibits the slowest performance, with a complexity of \(O(NL^2)\), as it performs a brute-force comparison of the query against all \(N\) audio sequences where L is the length of the audio sequence. Brute-force (TF-IDF) retrieval, with \(O(N)\) complexity, is faster but still inefficient for large-scale retrieval. The inverted index method, that leverages precomputed token-based indexing, significantly reduces search time, achieving a complexity close to \(O(\log N)\) by quickly retrieving relevant candidates without expensive similarity computations. H-QuEST, which combines HNSW-based retrieval with Smith-Waterman alignment, has a total complexity of \(O(\log N) + O(KL^2)\), where \(K \ll N\), making it much faster than DTW while maintaining high accuracy.

Table~\ref{tab:hs_comparison} presents a comparison of the False Rejection Rate (FRR) on the Hey-Snips (HS) dataset under clean and babble noise conditions across different Query-by-Example Keyword Spotting (QbE-KWS) approaches and H-QuEST. The baseline models (Small and Large) exhibit relatively higher FRR values, indicating lower performance in keyword spotting. The QbE-KWS models, which leverage multi-head attention and soft-triple loss, show significant improvements, reducing the FRR compared to the baseline. Among all approaches, the proposed H-QuEST method achieves the lowest FRR, with 6.30\% for HS (Clean) and 6.20\% for HS (Babble), demonstrating its superior robustness and accuracy in QbE-STD even in noisy conditions.

\section{Conclusion}
This paper introduces H-QuEST an effective QbE-STD system that enhances search efficiency through the application of HNSW-based indexing and Smith-Waterman refinement on TF-IDF vectors derived from discrete audio tokens generated using Wav2Vec2.0. Extensive experiments highlight the H-QuEST's superior performance compared to TF-IDF and attention based methods on several datasets. Results also show that H-QuEST improves spoken term detection accuracy (MAP) and significantly reduces time complexity in search compared to DTW, offering an efficient solution for navigating large audio datasets with precise retrieval.
\section{Acknowledgements}
This work was supported by a research grant from MeitY, Govt. of India, under the project BHASHINI.

\bibliographystyle{IEEEtran}
\bibliography{mybib}

\end{document}